\documentclass[twoside]{article}
\usepackage{fleqn,espcrc2}

\usepackage{graphicx}

\newcommand{\epem}   {\ensuremath{\mathrm{e^+e^-}}}
\newcommand{\wpwm}   {\ensuremath{\mathrm{W^+W^-}}}
\newcommand{\mpmm}   {\ensuremath{\mu^+\mu^-}}
\newcommand{\tptm}   {\ensuremath{\tau^+\tau^-}}
\newcommand{\qqbar}  {\ensuremath{\mathrm{q\overline{q}}}}
\newcommand{\bbbar}  {\ensuremath{\mathrm{b\overline{b}}}}
\newcommand{\bt}     {\ensuremath{B_T}}
\newcommand{\bw}     {\ensuremath{B_W}}

\newcommand{\thr}    {\ensuremath{1-T}}
\newcommand{\cp}     {\ensuremath{C}}
\newcommand{\bcf}    {\ensuremath{\overline{C}_F}}
\newcommand{\bca}    {\ensuremath{\overline{C}_A}}
\newcommand{\cf}     {\ensuremath{C_F}}
\newcommand{\ca}     {\ensuremath{C_A}}
\newcommand{\tf}     {\ensuremath{T_F}}
\newcommand{\nf}     {\ensuremath{n_f}}

\newcommand{\order}[1]{\ensuremath{\mathcal{O}(#1)}}
\newcommand{\xmu}    {\ensuremath{x_{\mu}}}
\newcommand{\as}     {\ensuremath{\alpha_{\mathrm{S}}}}
\newcommand{\asq}    {\ensuremath{\as(Q)}}

\newcommand{\oaa}    {\ensuremath{\mathcal{O}(\as^2)}}
\newcommand{\oaaa}   {\ensuremath{\mathcal{O}(\as^3)}}
\newcommand{\asmu}   {\ensuremath{\as(\mu)}}
\newcommand{\roots}  {\ensuremath{\sqrt{s}}}
\newcommand{\rootsp} {\ensuremath{\sqrt{s'}}}
\newcommand{\znull}  {\ensuremath{\mathrm{Z^0}}}
\newcommand{\mz}     {\ensuremath{M_{\znull}}}
\newcommand{\asmz}   {\ensuremath{\as(\mz)}}
\newcommand{\ycut}   {\ensuremath{y_{\mathrm{cut}}}}
\newcommand{\lmqcd}  {\ensuremath{\Lambda_{\mathrm{QCD}}}}
\newcommand{\mw}     {\ensuremath{M_{\mathrm{W}}}}
\newcommand{\stat}   {\ensuremath{\mathrm{(stat.)}}}
\newcommand{\syst}   {\ensuremath{\mathrm{(syst.)}}}

\newcommand{\theo}   {\ensuremath{\mathrm{(theo.)}}}
\newcommand{\chisq}  {\ensuremath{\chi^2}}
\newcommand{\chisqd} {\ensuremath{\chi^2/\mathrm{d.o.f.}}}
\newcommand{\lnr}    {\ensuremath{\ln(\mathrm{R})}}
\newcommand{\bthree} {\ensuremath{\mathrm{B}_3}}
\newcommand{\rthree} {\ensuremath{\mathrm{R}_3}}
\newcommand{\rthreeb}{\ensuremath{\mathrm{R^b}_3}}
\newcommand{\rthreel}{\ensuremath{\mathrm{R^l}_3}}
\newcommand{\rfour}  {\ensuremath{\mathrm{R}_4}}
\newcommand{\mb}     {\ensuremath{m_{\mathrm{b}}}}
\newcommand{\mbmz}   {\ensuremath{\mb(\mz)}}
\newcommand{\mbmb}   {\ensuremath{\mb(\mb)}}
\newcommand{\msbar}  {\ensuremath{\overline{\mathrm{MS}}}}
\newcommand{\chibz}  {\ensuremath{\chi_{\mathrm{BZ}}}}
\newcommand{\degr}   {\ensuremath{^{\mathrm{o}}}}
\newcommand{\kaph}   {\ensuremath{\kappa_{\mathrm{H}}}}
\newcommand{\ejet}   {\ensuremath{E_{\mathrm{jet}}}}
\newcommand{\xe}     {\ensuremath{x_{\mathrm{E}}}}
\newcommand{\ddel}   {\ensuremath{\mathrm{d}}}
\newcommand{\nchgg}  {\ensuremath{N^{\mathrm{ch.}}_{\mathrm{gg}}}}
\newcommand{\nchqq}  {\ensuremath{N^{\mathrm{ch.}}_{\qqbar}}}
\newcommand{\nchincl}{\ensuremath{N^{\mathrm{ch.}}_{\mathrm{incl.}}}}
\newcommand{\ktlu}   {\ensuremath{k_{\perp,\mathrm{Lu}}}}
\newcommand{\sqg}    {\ensuremath{s_{\mathrm{qg}}}}
\newcommand{\sqbarg} {\ensuremath{s_{\mathrm{\overline{q}g}}}}

\title{ Jet physics in \epem\ annihilation from 14 to 209~GeV }

\author{ S. Kluth\address{ Max-Planck-Institut f\"ur Physik,\\
         F\"ohringer Ring 6, D-80805 Munich, Germany } }

\begin{document}

\begin{abstract}
Results from the study of jets in hadronic final states produced in
\epem\ annihilation at $\roots=14$ to 209~GeV are discussed.  Precise
measurements of the strong coupling \as\ at many points of \roots\
provide convincing evidence for the running of \as\ as expected by
QCD.  Several measurements of the running b-quark mass \mbmz\ are
combined yielding $\mbmz=(2.92\pm0.03\stat\pm0.31\syst)$~GeV and
compared with low energy results.  Strong evidence for the running of
the b-quark mass is found.  Experimental investigations of the gauge
structure of QCD are reviewed and then combined with the results for
the colour factors $\bca=2.89\pm0.03\stat\pm0.21\syst$ and
$\bcf=1.30\pm0.01\stat\pm0.09\syst$ and correlation coefficient
$\rho=0.82$.  The uncertainties on the colour factors are below 10\%
and the agreement with the QCD expectation $\ca=3$ and $\cf=4/3$ is
excellent.
\vspace{1pc}
\end{abstract}

\maketitle

\section{ INTRODUCTION }

Hadron production in \epem\ annihilation is an ideal laboratory for
QCD studies.  For hadronic final states there is no interference with
the leptonic intial state.  The energy of the hadronic final state is
maximal in the laboratory system (in the case of symmetric colliders)
thus allowing efficient production and experimental study of highly
energetic hadronic systems.  Since the particles involved in the
electroweak production of hadrons via the process
$\epem\rightarrow (\znull\gamma)^*\rightarrow\qqbar$ are pointlike
there are no parton density functions to take into account.  

\begin{figure}[!htb]
\includegraphics[width=\columnwidth]{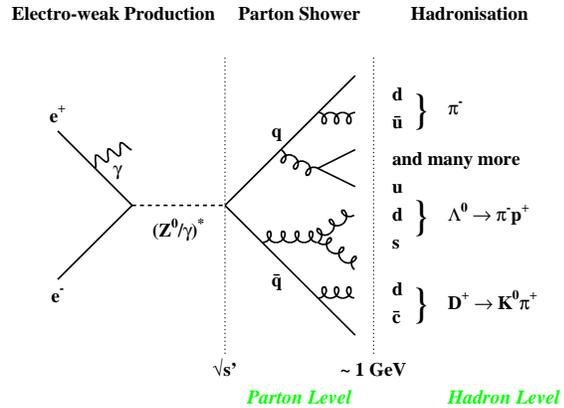}
\caption{ Schematic view of hadron production in \epem\ annihilation. }
\label{fig_qcdshower}
\end{figure}

The current understanding of hadron production in \epem\ annihilation
in the Standard Model is shown schematically in
figure~\ref{fig_qcdshower}.  The colliding electron and positron
annihilate into a photon or \znull, possibly with additional radiation
of energetic photons from the incoming beam particles (initial state
radiation or ISR).  The intermediate photon is always virtual with its
invariant mass equal to the invariant mass \rootsp\ of the \epem\
system at collision after ISR while the \znull\ can be produced on
mass-shell when $\rootsp\simeq\mz$.  The intermediate photon or
\znull\ then decays into a \qqbar\ state which can in turn radiate
gluons.  The continuing processes of gluon radiation and \qqbar\
production result in a parton shower which only terminates when the
partons (quarks and gluons) reach energy scales of $\sim 1$~GeV.  
Final states at this stage are referred to as parton-level.  At these
scales the increasing strong coupling causes the formation of
colourless bound states of quarks, i.e. hadrons.  This process,
known as hadronisation, is not well understood at a fundamental
level in QCD but many successful models based on QCD have been
developed.  The hadronic final state after decays of short lived
particles (usually defined to have lifetimes $<300$~ps) is said as
being at the hadron-level.

It is well known that the hadronic final states in \epem\ annihilation
show a so-called jet structure, i.e. several groups of particles are
observed in relatively small and separate regions of the solid angle
around the interaction point.  These structures are interpreted as
remnants of the \qqbar\ pair from the virtual photon or \znull\ decay
and possible energetic gluon radiation.  Many definitions of jet
finding algorithms~\cite{bethke92,dokshitzer97b} or event shape
observables (see e.g. ~\cite{OPALPR054,OPALPR075}) are used to
quantify the jet structure and compare with predictions from QCD.

Jet finding algorithms cluster particles lying closely together in
phase space.  As an example in the Durham algorithm~\cite{durham} the
distance between two particles $i$ and $j$ with energies $E_i$ and
$E_j$ and spanning the angle $\theta_{ij}$ is defined as
$y_{ij}=2E_iE_j(1-\cos\theta_{ij})/s$.  The pair of particles with the
smallest $y_{ij}$ is combined into a pseudo-particle by adding the
four-momenta and is then discarded.  The procedure is iterated until
only one jet remains.  One can now study the jet structure as a
function of the resolution parameter \ycut\ and e.g. determine the
fraction of 3-jet events at a fixed value of \ycut.  Alternatively,
the value of $y_{ij}=y_{32}$ where an event changes from a 3-jet to a
2-jet configuration is characteristic for its structure, i.e. a large
value of $y_{32}$ indicates the presence of a significant third jet.

At the time of writing data of observables such as jet production
rates or event shape distributions measured at centre-of-mass energies
\roots\ ranging from 14 to 209~GeV are available.  These are especially
interesting, because many QCD predictions are strongly energy
dependent due to the running of the strong coupling \asq.  In leading
order the running of \as\ is expressed as
\begin{equation}
\label{equ_asrun}
  \asq = \asmu/(1+2\beta_0\asmu\ln(\mu/Q))
\end{equation}
with $\beta_0=(11\ca-2\nf)/(12\pi)$.  The value \asq\ at a scale
$Q$ is given by \asmu\ at a reference scale $\mu$ and the logarithmic
ratio of the energy scales $2\ln(Q/\mu)$.  However, hadronisation
effects scale like $1/Q^n$, with $n=1$ for many jet algorithms or
event shape observables, implying that a meaningful comparison of QCD
predictions with data requires a detailed understanding of the scale
dependencies of perturbative QCD and as well as of the hadronisation
effects.  The present \epem\ data allow to study these problems in a
controlled way.

\begin{figure}[!htb]
\includegraphics*[width=\columnwidth]{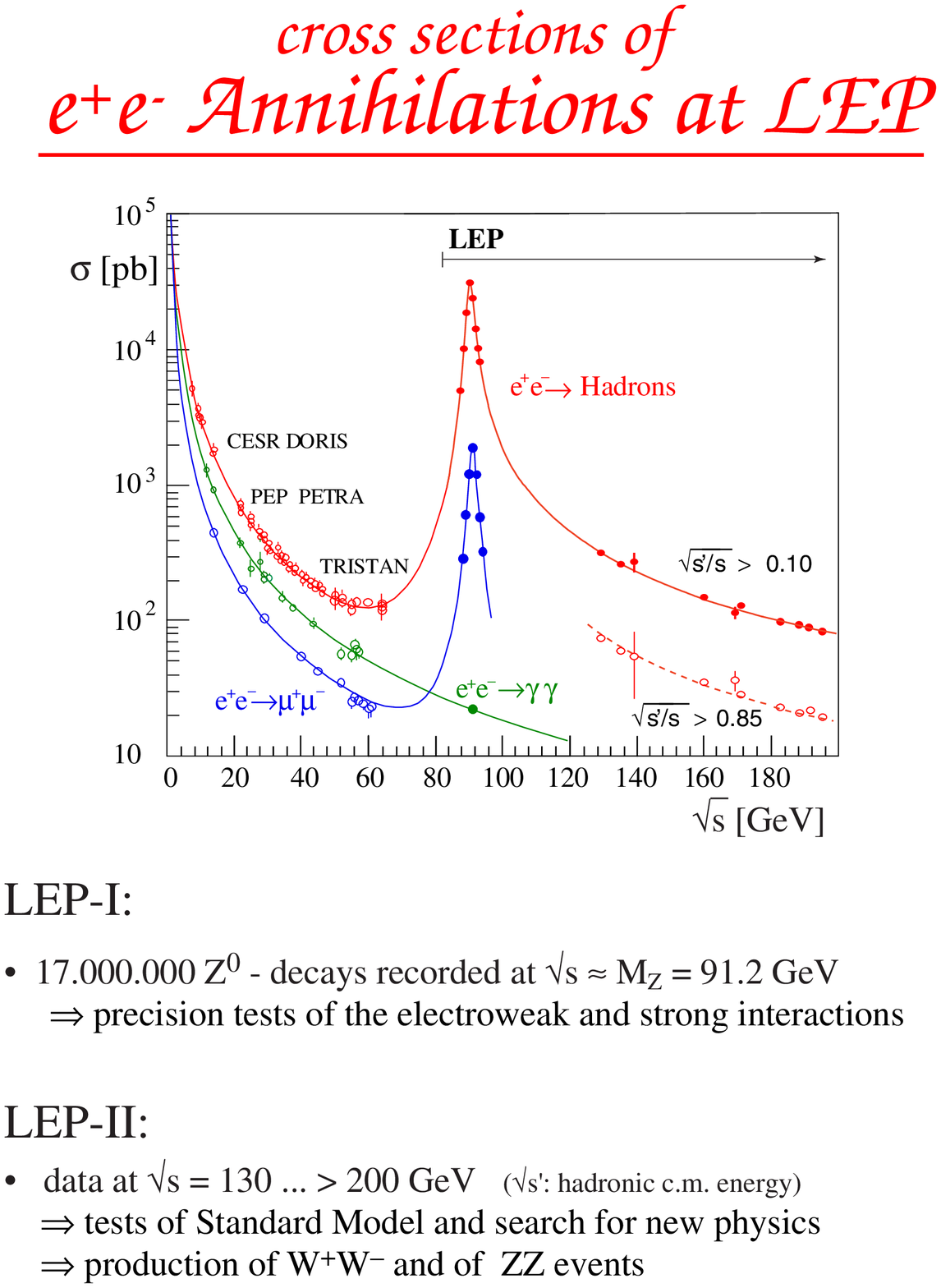}
\caption{ Cross sections for various \epem\ annihilation processes as
indicated on the figure~\cite{bethke00b}. }
\label{fig_wq}
\end{figure}

The selection of \epem\ annihilation into hadronic final states has to
consider different experimental conditions at the various cms
energies.  Figure~\ref{fig_wq} shows the total cross sections for
hadron production, \mpmm\ pair production and two-photon
interactions~\cite{bethke00b}.  At $\roots<\mz$ backgrounds stem from
two-photon processes, \tptm\ pair production and events with hard ISR.
For example in the JADE data recorded between $\roots=14$ and 44~GeV
after demanding large total energy, balance of momentum along the beam
axis and large particle multiplicity the total background is reduced
to about 1\% with a selection efficiency of about
60\%~\cite{pedrophd}.  On the
\znull\ peak similar event selections reduce the background to less
than 1\% with almost 100\% selection efficiency.  Above the \znull\
peak for $\roots>2\mw$ backgrounds from hadron production mainly via
production of \wpwm\ pairs decaying both hadronically become
important.  The prominent 4-jet structure of the \wpwm\ events and
their different kinematic properties compared to
$\epem\rightarrow\qqbar$ events allows to reduce backgrounds to less
than 10\% at selection efficiencies of about 80\%.

\section{ HARD QCD }

\subsection { Measurements of \as }

The strong coupling is in QCD predictions for massless partons the
only free parameter.  A fit of the theoretical prediction to the data
with \as\ as a free parameter yields a measurement of \asq\ where $Q$
is identified with the hard scale of the process, i.e. $Q=\roots$.
Hadronisation effects are not part of the perturbative QCD prediction
and are usually treated as a correction which is applied to the theory
before comparison with the data.  The choice between several
hadronisation models implemented in Monte Carlo generator programs and
uncertainties in the experimentally adjusted free parameters of the
models are reflected in the systematic uncertainties of the \as\
determinations.

\begin{figure}[!htb]
\includegraphics*[width=\columnwidth]{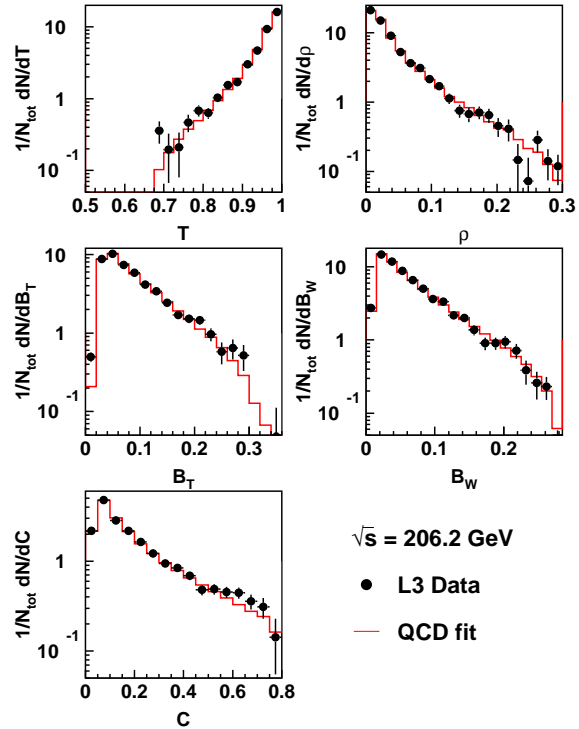}
\caption{ Data from L3 of event shape observables measured at
$\roots=206.2$~GeV. Superimposed are \oaa+NLLA QCD
fits~\cite{l3254}. } 
\label{fig_l3as}
\end{figure}

Figure~\ref{fig_l3as} shows the measurement of
$\as(206.2~\mathrm{GeV})$ by L3~\cite{l3254} using a set of five event
shape observables Thrust $T$, Heavy Jet Mass $\rho$, Total and Wide
Jet Broadening \bt\ and \bw, and C-parameter $C$.  For these
observables matched \oaa+NLLA QCD
calculations~\cite{nllathmh,nllabtbw2,nllacp} are available, which
provide reliable predictions over a large part of the phase space
populated by 2- and 3-jet events.  The combined result of the five
measurements is
$\as(206.2~\mathrm{GeV})=0.113\pm0.002\stat\pm0.005\syst$.  

\begin{figure}[!htb]
\includegraphics*[width=\columnwidth]{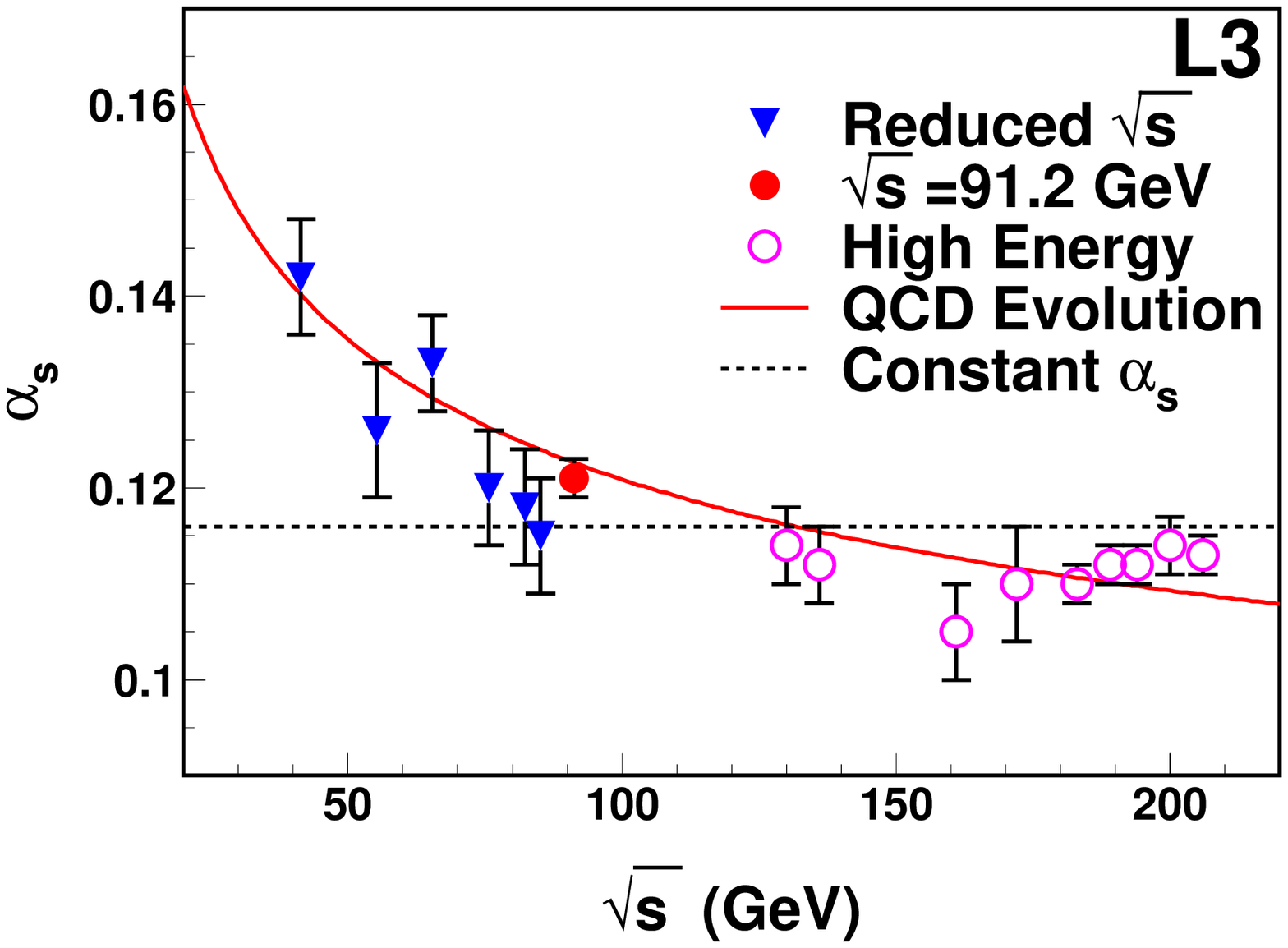}
\caption{ Results of $\as(\roots)$ from L3 compared with the
expectation from QCD for the running of \as\ (solid line) and with
$\as=\mathrm{const.}$ (dashed line)~\cite{l3254}. }
\label{fig_l3asrun}
\end{figure}

Figure~\ref{fig_l3asrun} presents the results of similar analysis
performed by L3 at the other LEP~2 energy points, at LEP~1 on the
\znull\ peak and using hadronic \znull\ decays with hard final state
photon radiation (FSR) for measurements at $\roots<\mz$.  The error
bars show the experimental and statistical uncertainties only.  The
solid line shows a fit of the QCD prediction for the running of \as\
with the result $\asmz=0.123\pm0.001\exp\pm0.006\theo$ and
$\chisqd=18/15$.  The assumption of \as=const. results in
$\chisqd=52/15$; the data thus support the running of \as\ as
predicted by QCD.  Similar analyses by the LEP collaborations
are~\cite{delphi210,OPALPR299}.  

The data of the former JADE experiment taken between 1979 and
1986 at PETRA/DESY at $\roots=14$ to 44~GeV have been re-analysed
recently~\cite{pedrophd}.  The analysis employs modern Monte Carlo
event generators together with the original detector simulation
program to obtain more reliable experimental corrections with reduced
statistical and systematic uncertainties.  The contribution of
$\epem\rightarrow\bbbar$ events is subtracted to reduce the effects of
heavy quark masses and heavy hadron decays on the analysis.  The data
can be compared directly with QCD predictions for massless partons
corrected for hadronisation effects.  Several event shape observables
are used to determine \as\ in the same way as the LEP collaborations
using the best currently available matched \oaa+NLLA QCD
calculations.

\begin{figure}[!htb]
\includegraphics*[width=\columnwidth]{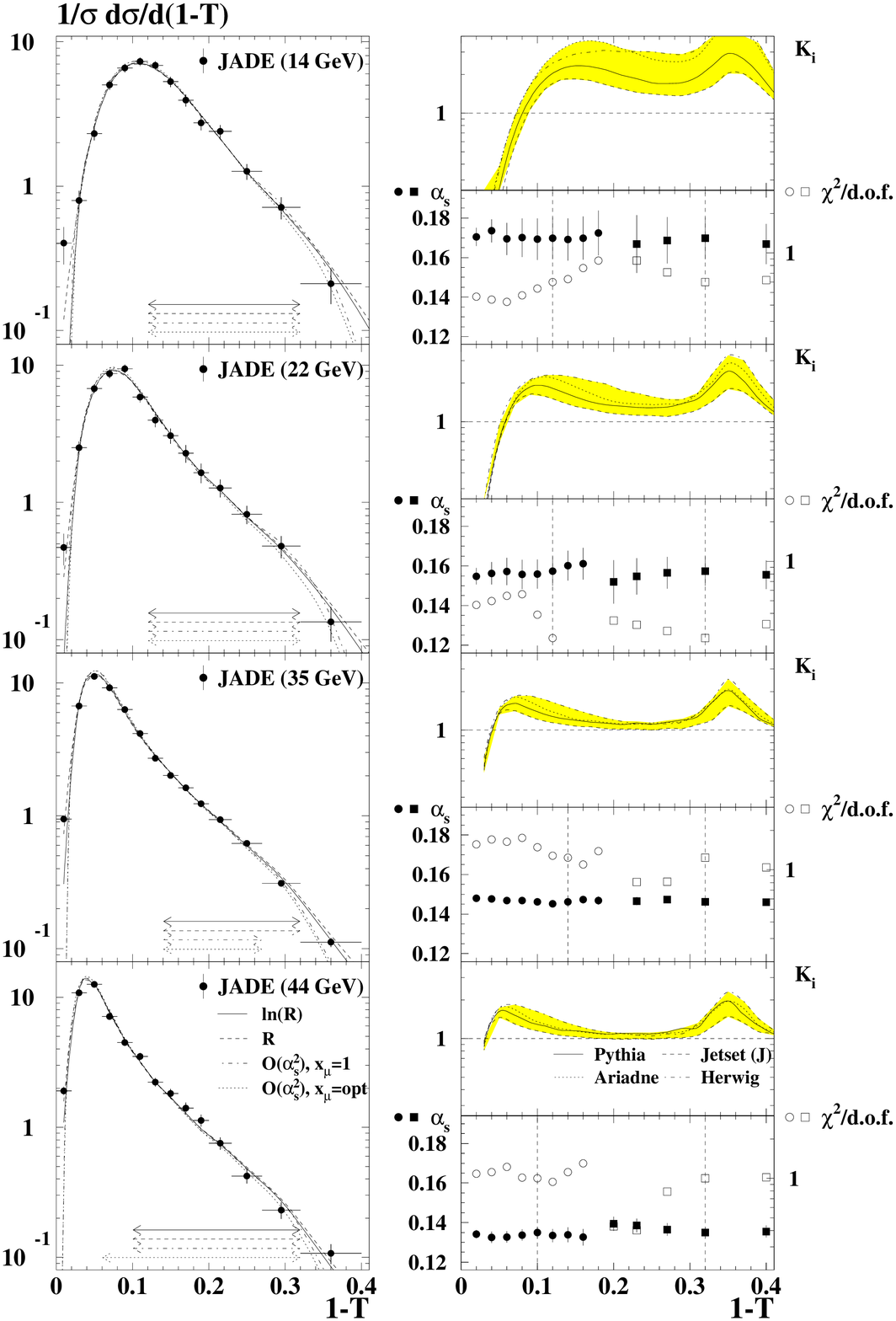}
\caption{ Data for \thr\ from JADE at $\roots=14$, 22, 35 and 44~GeV
are shown on the left with \oaa+NLLA or \oaa\ QCD fits superimposed
(lines).  On the right the hadronisation corrections $K_i$ are shown
with uncertainties as grey bands~\cite{pedrophd}.  } 
\label{fig_jadeth}
\end{figure}

The results of fits to \thr\ are shown in figure~\ref{fig_jadeth}.
The diagrams show the data for \thr\ at $\roots=14$, 22, 35 and 44~GeV
corrected for experimental effects.  A decrease of the shoulder at
large \thr\ with increasing \roots\ is clearly visible; this
observation corresponds to the expected lower fraction of events with
hard gluon radiation due to the running of \as.  Superimposed are the
results of QCD fits using matched \oaa+NLLA (\lnr- and R-matching) or
\oaa\ calculations corrected for hadronisation effects, which
generally describe the data well.  The hadronisation corrections $K_i$
with uncertainties are shown as grey bands.  The hadronisation
corrections are large with large uncertainties at low \roots\ but
become much smaller with increasing \roots.  The results for
$\as(\roots)$ are 
$\as(14~\mathrm{GeV})=0.170\pm0.019$,
$\as(22~\mathrm{GeV})=0.151\pm0.013$,
$\as(35~\mathrm{GeV})=0.146\pm0.012$ and
$\as(44~\mathrm{GeV})=0.131\pm0.009$. 
The data together with other compareable results from the LEP
experiments and SLD are in good agreement within the experimental
errors with the QCD prediction for the running 
\as\ based on $\asmz=0.118\pm0.003$~\cite{bethke00a}.  As above a
fit assuming \as=const. results in a \chisq\ probability of $\sim
10^{-5}$.  

A novel method to determine the strong coupling \as\ uses the rate of
4-jet events \rfour\ determined using the Durham
algorithm~\cite{aleph249,delphiconf487}.  Recently the NLO corrections for
observables sensitive to at least four partons in the final state,
i.e. corrections of \oaaa, became available~\cite{dixon97,nagy98b}.
These calculations are matched with existing NLLA calculations.

\begin{figure}[!htb]
\includegraphics*[width=\columnwidth]{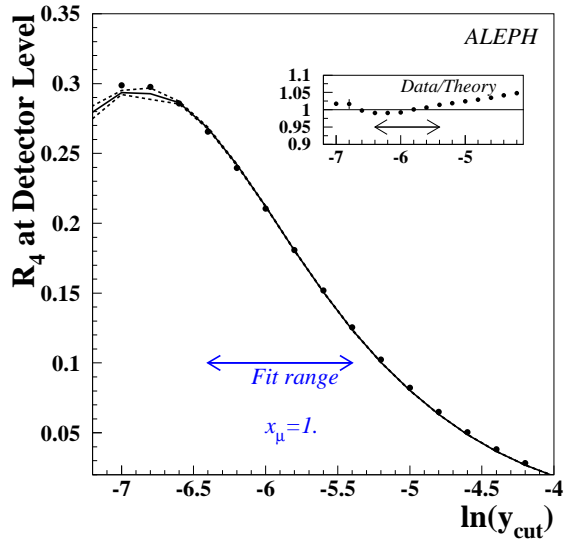}
\caption{ The 4-jet fraction \rfour\ determined using the Durham
algorithm is shown as a function of \ycut\ without corrections.
Superimposed is a fit of \oaaa+NLLA QCD (solid
line)~\cite{aleph249}. } 
\label{fig_aleph4jet}
\end{figure}

A measurement of \asmz\ based on these predictions is presented
in figure~\ref{fig_aleph4jet}~\cite{aleph249}.  The uncorrected data
for \rfour\ measured using $2.5\cdot 10^6$ hadronic \znull\ decays
using the Durham algorithm are shown as a function of the jet
resolution parameter \ycut.  The line shows the fit of the
\oaaa+NLLA (R-matching) theory corrected for hadronisation and detector
effects.  The fit is performed using only the bins within the fit
range as indicated where the total correction factors deviate from
unity by less than 10\%.  The result of the fit for fixed
renormalisation scale parameter $\xmu=1$ is $\asmz=0.1170\pm0.0022$
using the conservatively estimated total error from~\cite{aleph249}.
The fit is seen to describe the precise data fairly well.  The similar
analysis of~\cite{delphiconf487} uses only the \oaaa\ calculations
with experimentally optimised renormalisation scale and finds
$\asmz=0.1178\pm0.0029$ with $\xmu^2=0.015$.  These measurements have
comparatively small uncertainties and belong to the group of the most
precise determinations of \as.  

\subsection{ Running b quark mass }

The QCD predictions discussed so far have been made for massless
partons.  The effects of massive quarks on the perturbative QCD
predictions are known in \oaa\ as
well~\cite{rodrigo97,bernreuther97,nason97}.  The mass of the heavy
quark \mb\ identified with the b quark at LEP energies becomes an
additional parameter of the theory and is subject to renormalisation
analogously to the strong coupling constant.  As a result the heavy
quark mass \mb\ will depend on the scale of the hard process in which
the heavy quark participates.  In leading order in the \msbar\
renormalisation scheme the running b-quark mass is
$\mb(Q)=M_{\mathrm{b}}(\as(Q)/\pi)^{12/23}$, where $M_{\mathrm{b}}$ is
the so-called pole mass defined by the pole of the renormalised heavy
quark propagator.  The running \mb\ is known to four-loop
accuracy~\cite{vermaseren97}.  Using low energy measurements of \mbmb\
one finds that $\mbmz\simeq 3$~GeV.

The mass of the b quark \mb\ is experimentally accessible in hadronic
\znull\ decays, because i) about 21\% of the decays are
$\znull\rightarrow\bbbar$ and these events can be identified with high
efficiency and purity, ii) event samples of $\order{10^6}$ events are
available and iii) observables like the 3-jet rate $\rthree(\ycut)$ in
the JADE or Durham algorithm can have enhanced mass effects going like
$\mb^2/\mz^2/\ycut$~\cite{bilenky95}.  In the experimental
analyses~\cite{delphi177,OPALPR336,brandenburg99,aleph223,delphiconf644}
the ratio $\bthree=\rthreeb/\rthreel$, \rthreeb (\rthreel) are the 3-jet
rates in b quark (light quark) events, is studied which reduces the
influence of common systematics.

\begin{figure}[!htb]
\includegraphics*[width=\columnwidth]{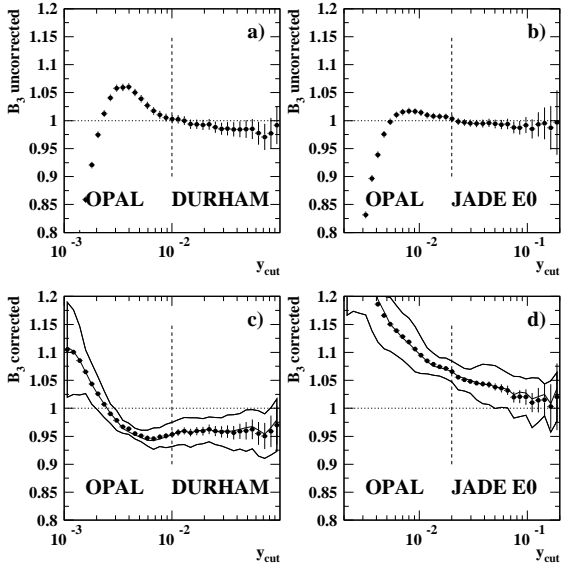}
\caption{ The figures show distributions of \bthree\ determined using
the Durham or JADE E0 jet finding algorithms before and after all
corrections. The solid lines in the lower row of figures give the
uncertainties on \bthree.  The vertical dashed lines indicate the
values of \ycut\ where the QCD predictions are
calculated~\cite{OPALPR336}. }
\label{fig_opalb3}
\end{figure}

Figure~\ref{fig_opalb3} shows $\bthree(\ycut)$ measured by
OPAL~\cite{OPALPR336} before and after detector and hadronisation
corrections for the JADE E0 and the Durham algorithm.  One observes
that the effect of the b quark mass is about 5\% on \bthree\ with
reasonable uncertainties as indicated by the bands.  The effect goes
in opposite directions for JADE E0 and the Durham algorithm, because
in the JADE type algorithms based on invariant mass to cluster jets
the large b quark mass causes an enhancement of \bthree\ while in the
Durham algorithm the reduced phase space for hard gluon emmission
dominates thus reducing
\bthree~\cite{brandenburg99}.  

\begin{figure}[!htb]
\includegraphics*[width=\columnwidth]{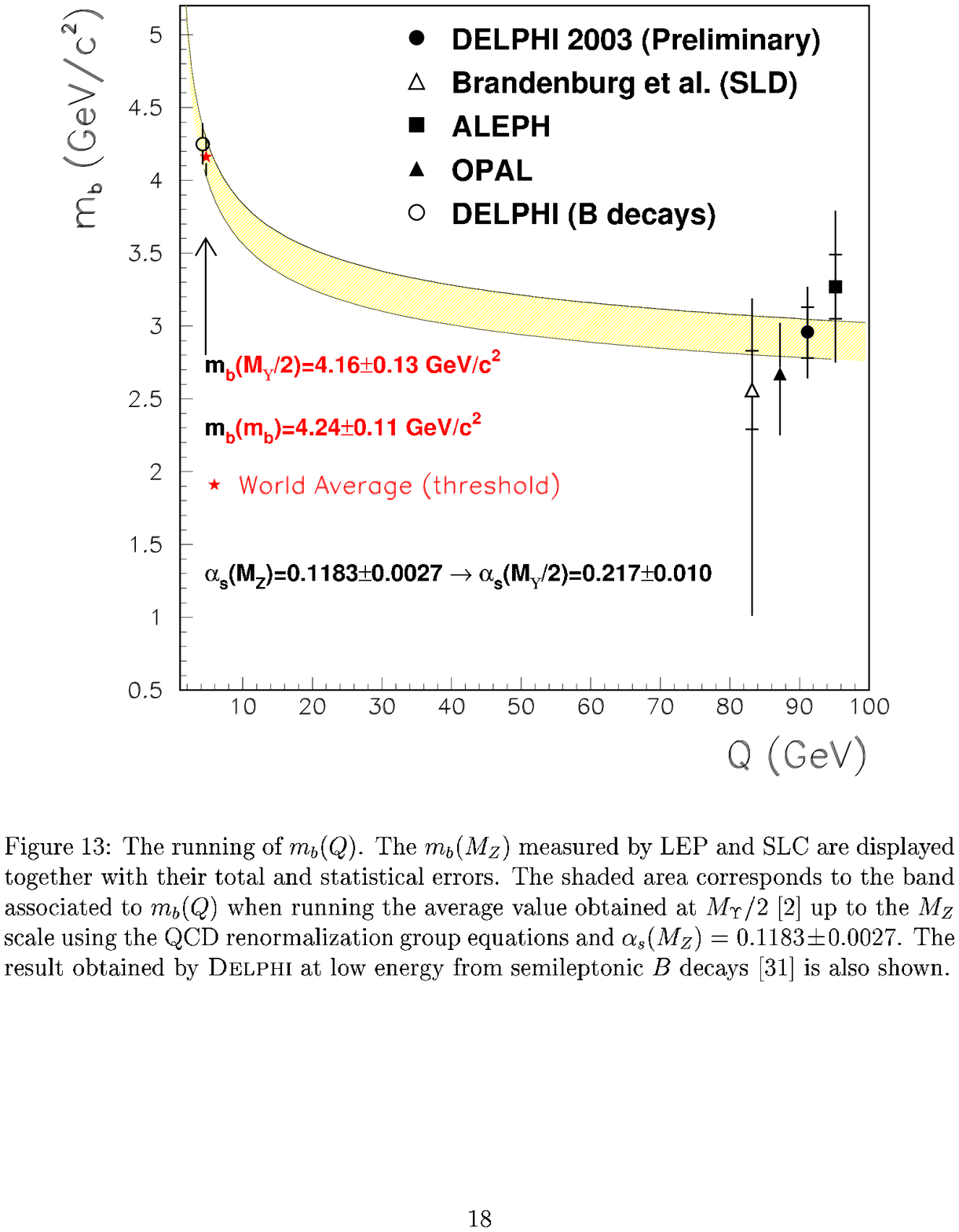}
\caption{ Measurements of $\mb(Q)$ at low and high energies $Q$ are
compared with the QCD expectation for the running of
$\mb(Q)$~\cite{delphiconf644}. }
\label{fig_delphimb}
\end{figure}

Figure~\ref{fig_delphimb} gives an overview over presently available
measurements of \mbmz~\cite{delphiconf644}.  These are compared with
low energy measurements and the QCD prediction for the running of
$\mb(Q)$.  The predicted range of \mbmz\ is in good agreement with the
measurements.  The measurements of \mbmz\
of~\cite{OPALPR336,brandenburg99,aleph223,delphiconf644} can be
averaged assuming the statistical errors to be uncorrelated, the
experimental and hadronisation uncertainties to be partially
correlated and the theory uncertainties to be fully
correlated\footnote{For partial correlation the covariance matrix
element is defined as $V_{i,j}=\min(\sigma_i,\sigma_j)^2$ and for full
correlation as $V_{i,j}=\sigma_i\sigma_j$}.  The result is
$\mbmz=(2.92\pm0.03\stat\pm0.31\syst)$~GeV, which may be
compared with the average of low energy measurements
$\mbmb=(4.24\pm0.11)$~GeV~\cite{el-khadra02}.  The difference becomes
$\mbmz-\mbmb=(1.32\pm0.33)$~GeV which is non-zero by four standard
deviations thus providing strong experimental evidence for the running
heavy quark mass in QCD.

\section{ QCD GAUGE STRUCTURE }

QCD as the gauge theory of strong interactions assumes that quarks
carry one out of three strong charges, referred to as colour.  The
requirement of local gauge symmetry under SU(3) transformations in the
colour space generates the gauge bosons of QCD: an oktett of gluons
each carrying colour charge and anti-charge, see e.g.~\cite{ellis96}.
The gluons can thus interact with themselves; it is due to this
property of the QCD gauge bosons that the theory explains confinement
and asymptotic freedom via the running of the strong coupling \as.  

In perturbative QCD at NLO three fundamental vertices contribute: i)
the quark-gluon vertex with colour factor \ca, ii) the gluon-gluon
vertex with colour factor \ca\ and iii) the \qqbar\ production from a
gluon with colour factor $\tf\nf$, see e.g.~\cite{magnoli90}.  The
colour factors are $\cf=4/3$, $\ca=3$ and $\tf\nf=1/2\cdot 5$ in QCD
with SU(3) gauge symmetry and specify the relative contribution of the
corresponding vertex to observables.  In NLO QCD the prediction for an
observable $R$ is $R= A\as +
(B_{\cf}\cf+B_{\ca}\ca+B_{\tf}\tf\nf)\cf\as^2$; NLLA predictions e.g. for
event shapes or jet rates decompose in a similar way.  For NLO
predictions for 4-jet observables an analogous decomposition in terms
of the six possible products of two out of the three colour factors
holds~\cite{nagy98a}.  

Experimental investigations of the gauge structure of QCD are possible
because of the different angular momenta in the initial and final
states of the fundamental vertices.  It is an important test of QCD to
probe the gauge structure in experiments.  Several techniques with
rather different experimental and theoretical uncertainties have been
developed; we will discuss here some recent results.

\subsection{ Four-jet events }

The LEP experiments ALEPH~\cite{aleph249} and OPAL~\cite{OPALPR330}
have analysed 4-jet final states from hadronic \znull\ decays using
the recent QCD NLO predictions (see e.g.~\cite{nagy98a} and references
therein).  The 4-jet final states are selected by clustering events
using the Durham algorithm~\cite{durham} with $\ycut=0.008$ and
demanding four jets.  At this value of \ycut\ the 4-jet fraction is
relatively large ($\rfour\simeq 7$\%) and the four jets are well
separated.  

The energy-ordered 4-momenta $p_i, i=1,\ldots, 4$ of the jets are used
to calculate the angular correlation observables (see
~\cite{aleph249,OPALPR330} for details).  As an example, the
Bengtsson-Zerwas angle \chibz\ is defined by
$\chibz=\angle([\vec{p_1}\times\vec{p_2}],[\vec{p_3}\times\vec{p_4}])$,
i.e. the angle between the two planes spanned by the momentum vector
pairs $(\vec{p_1},\vec{p_2})$ and $(\vec{p_3},\vec{p_4})$.  Assuming
that energy ordering selected the primary quarks from the \znull\
decay as $(p_1,p_2)$ the observable \chibz\ is sensitive to the decay
of an intermediate gluon to gluons (vertex ii)) or quarks (vertex
iii)) and the competing process of radiation of a second gluon from a
primary quark (vertex i)).

\begin{figure}[!htb]
\includegraphics*[width=\columnwidth]{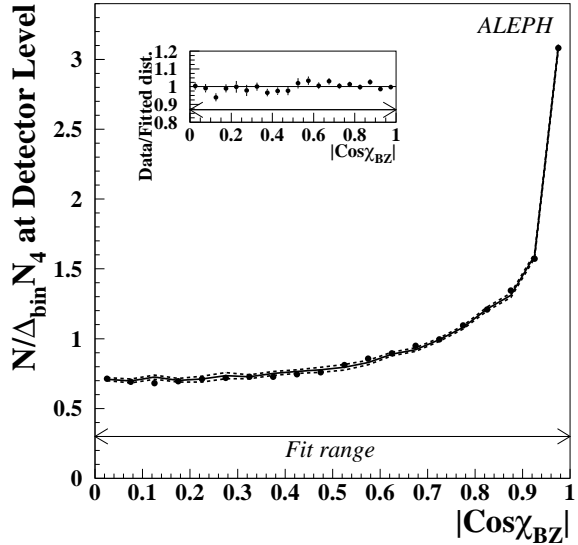}
\caption{ The distribution of $|\cos\chibz|$ before corrections is
shown (solid points).  Superimposed is a fit of NLO QCD (solid
lines)~\cite{aleph249}.  }
\label{fig_cosbz}
\end{figure}

Figure~\ref{fig_cosbz} presents the uncorrected distribution of
$\cos(\chibz)$ measured by ALEPH~\cite{aleph249}.  Superimposed on the
data points is the result of a simultaneous fit of the NLO QCD
predictions to four angular correlations including \chibz\ and the
distribution of \rfour.  From the fit values for \asmz\ and the colour
factors \ca\ and \cf\ are extracted.  The results from ALEPH are
$\asmz=0.119\pm0.027$, $\ca=2.93\pm0.60$ and $\cf=1.35\pm0.27$; from
OPAL we have $\asmz=0.120\pm0.023$, $\ca=3.02\pm0.55$ and
$\cf=1.34\pm0.26$.  The systematic errors are dominated by
uncertainties from the hadronisation corrections and theoretical
errors from estimating missing higher order
contributions\footnote{Both analyses use an unconventional method of
estimating systematic uncertainties.  A more conservative evaluation
leads to total errors for \asmz, \ca\ and \cf\ of $\pm0.048$,
$\pm1.06$ and $\pm0.46$ for ALEPH and $\pm0.049$, $\pm1.07$ and
$\pm0.47$ for OPAL.}.

\subsection{ Event shape fits }

In this analysis~\cite{colrun} the decomposition of the \oaa+NLLA QCD
predictions for event shape observables into terms proportional to the
colour factors is used.  Since the sensitivity of event shape
distributions measured at LEP~1 alone is not
sufficient~\cite{OPALPR134} data from $\roots=14$ to 189 GeV are used.
In this way the colour structure of the running of the strong coupling
contributes as well.  Hadronisation corrections are implemented using
analytic predictions known as power corrections (see~\cite{colrun} and
references therein).  The power corrections essentially predict that
hadronisation effects cause a shift of the perturbative distribution
proportial to $1/\roots$ in the region where the \oaa+NLLA predictions
are reliable.  The advantage of using power corrections instead of
Monte Carlo model based hadronisation corrections is that the colour
structure of the power corrections is known and can be varied in the
fit. 

In the analysis simultaneous fits of \asmz, \ca\ and \cf\ to data for
the event shape observables \thr\ at $\roots=14$ to 189~GeV and \cp\
at $\roots=35$ to 189~GeV are performed.  The data for \thr\ and \cp\
are analysed separatly and the results are combined.  The results are
$\asmz=0.119\pm0.010$, $\ca=2.84\pm0.24$ and $\cf=1.29\pm0.18$ and are
shown on figure~\ref{fig_cacf} below.  The errors are dominated by
uncertainties from the hadronisation correction and from experimental
effects.

\subsection{ Scaling violation in gluon and quark jets }

Jets originating from quarks or gluons should have different
properties due to the different colour charge carried by quarks or
gluons~\cite{brodsky76,konishi78}.  In an analysis by
DELPHI~\cite{delphi236} the scaling violation of the fragmentation
function (FF) in gluon and quark jets at different energies is
compared.  From a sample of $3.7\cdot 10^6$ hadronic \znull\ decays
planar 3-jet events with well reconstructed jets are selected using
the Durham or Cambridge algorithms, but without imposing a fixed value
of \ycut\ in order to reduce biases~\cite{eden98}.  In events with the
two angles between the most energetic jet and the other jets in the
range between 100\degr\ and 170\degr\ a b-tagging procedure is applied
to the jets.  Gluon jets are identified indirectly as the jets without a
successful b-tag.  Jets originating from light (udsc) quarks are
taken from events which failed the b-tagging.  A correction procedure
based on efficiencies and purities determined by Monte Carlo
simulation yields results for pure gluon or quark jets.  

\begin{figure}[!htb]
\includegraphics*[width=\columnwidth]{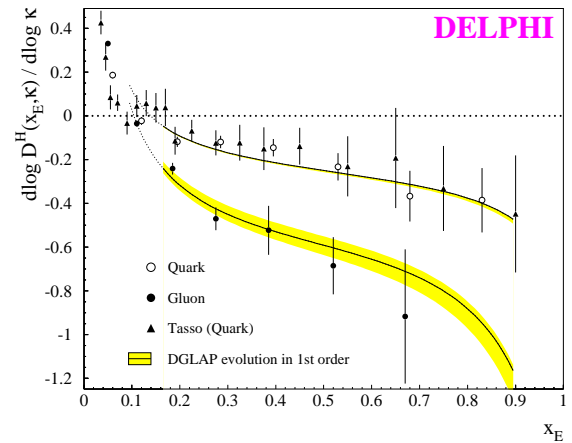}
\caption{ Measurements of $\ddel D^H(\xe,\kaph)/\ddel\log(\kaph)$ are
shown for gluon (solid points) and quark jets (open circles and solid
triangles).  The lines represent a fit of the LO QCD
prediction~\cite{delphi236}.  } 
\label{fig_delphisvgq}
\end{figure}

The jet energy scale is calculated according to
$\kaph=\ejet\sin(\theta/2)$, where $\theta$ is the angle w.r.t. to the
closest jet.  This definition takes colour coherence effects into
account~\cite{basicspqcd}.  The FF in a jet $D^H(\xe,\kaph)$ is given
by the distribution of $\xe=E_{\mathrm{hadron}}/\kaph$ for the hadrons
assigned to the jet.  The evolution of the FFs of gluon and quark jets
with jet energy scale \kaph\ is studied in intervals of \xe.
Figure~\ref{fig_delphisvgq} presents the quantity $\ddel
D^H(\xe,\kaph)/\ddel\log(\kaph)$, i.e. the slopes of the scaling
violation for a given interval of \xe, for quark and gluon jets.  The
steeper slopes corresponding to stronger scaling violations of gluon
compared to quark jets is clearly visible.  A fit of the scaling
violations based on the LO QCD prediction (DGLAP equation) allows to
extract the ratio $\ca/\cf$ resulting in $\ca/\cf=2.26\pm0.16$.

\subsection{ Multiplicity in unbiased gluon jets }

In this analysis~\cite{OPALPR347} the evolution with energy scale of
the charged particle multiplicities \nchgg\ and \nchqq\ for gluon and
quark jets is used to measure the colour charge ratio $\ca/\cf$.  The
inclusive multiplicity \nchincl\ is measured in hadronic \znull\
decays as a function of the 3-jet transverse momentum scale
$\ktlu=\ln(\sqg\sqbarg/(s\lmqcd)$ where \sqg\ and \sqbarg\ are the
invariant masses between the gluon and the quark
jets~\cite{eden98,eden99}.  The analysis reconstructs three jets in
all events by adjusting the value of \ycut\ in the Durham algorithm,
selects one-fold symmetric Y-events~\cite{OPALPR038} and identifies
the lowest energy jet as the gluon jet.  By comparing with
measurements of $\nchqq(\ktlu)$ the multiplicity in unbiased gluon
jets \nchgg\ can be extracted as a function of the jet energy
scale \ktlu: $\nchgg(\ktlu)=2(\nchincl(\ktlu)-\nchqq(\ktlu))$.  

\begin{figure}[!htb]
\includegraphics[clip=true,width=\columnwidth]{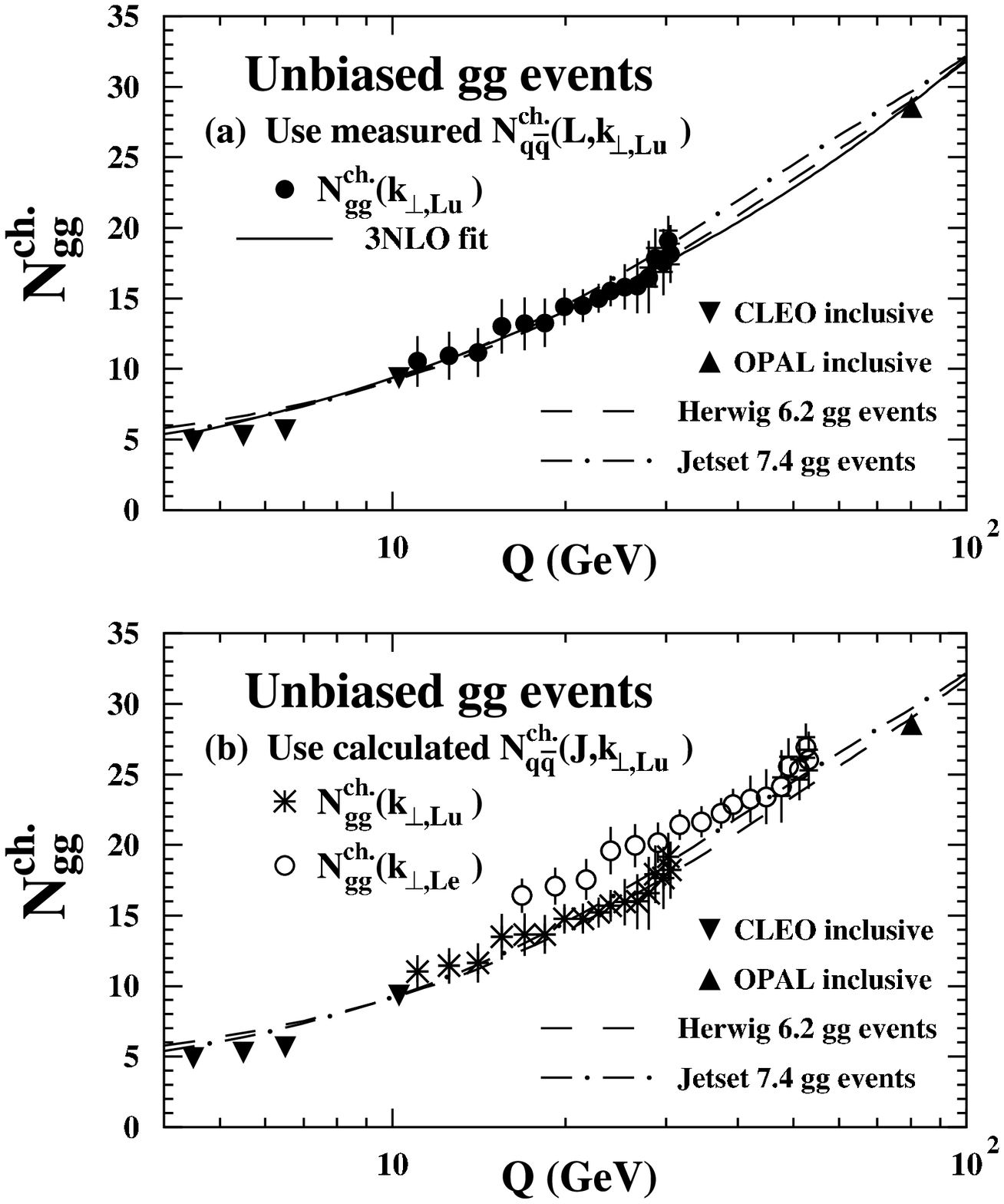}
\caption{ Measurements of $\nchgg(Q)$ are shown using 3-jet events
(solid points) and other data (solid triangles).  Superimposed are
predictions of Monte Carlo generators (dashed and dash-dotted lines)
and of the QCD fit (solid line)~\cite{OPALPR347}. }
\label{fig_opalpr347_06a}
\end{figure}

Figure~\ref{fig_opalpr347_06a} shows the data for \nchgg\ together
with compareable data at lower and higher jet energy scales \ktlu\
from CLEO and OPAL.  The solid line represents the result of a fit of
the QCD prediction for the evolution of \nchgg\ compared to
\nchqq~\cite{eden98}.  The fit with $\ca/\cf$ as a free parameter
yields $\ca/\cf=2.23\pm0.14$.  This analysis is complementary to the
analysis of scaling violations discussed above because the
multiplicity is dominated by low energy particles while the scaling
violation is dominated by high energy particles with large
\xe.

\subsection{ Colour factor averages }

The measurements of the colour factors \ca\ and \cf\ or of $\ca/\cf=x$
discussed above can be combined into average values of \ca\ and \cf\
taking into account correlations between \ca\ and \cf\ as well as
between different experiments.  The variables $x$ and $y=\tf/\cf$ are
used to define the \chisq\ function
\begin{eqnarray}
\label{equ_covxy}
 \chisq & = & \sum_i (x_i-\bar{x})v^{(xy)}_{i}(y_i-\bar{y}) + \\ \nonumber
 & &        \sum_{ij} (x_i-\bar{x})v^{(x)}_{ij}(x_j-\bar{x}) + \\ \nonumber
 & &        \sum_{ij} (y_i-\bar{y})v^{(y)}_{ij}(y_j-\bar{y})\;\;, \\ \nonumber
\end{eqnarray}
where $\bar{x}$ and $\bar{y}$ are the averages, the indices $i,j$
count experiments, $v^{(xy)}_{i}$, $v^{(x)}_{ij}$ and $v^{(y)}_{ij}$
are elements of the inverses of the corresponding covariance matrices
for the $x_i$ and $y_i$ within experiment $i$ and for the $x_i$ and
$y_i$ between experiments.  The averages $\bar{x}$ and $\bar{y}$
are converted to the average values \bca\ and \bcf\ after the fit is
performed.

The input data for $x$ and $y$ are directly taken
from~\cite{aleph249,OPALPR330,delphi236,OPALPR347} while the results
from~\cite{colrun} have to be converted\footnote{The results are
$x=2.20\pm0.21\stat\pm0.25\syst$, $y=0.388\pm0.021\stat\pm0.051\syst$,
$\rho_{\mathrm{stat.}}=0.98$ and $\rho_{\mathrm{syst.}}=0.86$.}.  

The covariance matrices between experiments are constructed as
follows: ALEPH and OPAL 4-jet analyses experimental and hadronisation
errors partially and theory errors fully correlated, 4-jet analyses
and event shape analysis hadronisation errors partially and theory
errors fully correlated, and 4-jet and event shape analyses and DELPHI
FF and OPAL \nchgg\ analyses theory errors partially correlated.

The averaging fit is done using only the first term of
equation~\ref{equ_covxy} to determine the averages \bca\ and \bcf,
using only statistical correlations in the first term to determine the
statistical errors and using the full covariance matrix to determine
the total errors.  The final results are
$\bca=2.89\pm0.03\stat\pm0.21\syst$,
$\bcf=1.30\pm0.01\stat\pm0.09\syst$ with correlation coefficient
$\rho=0.82$.  The relative total uncertainties are about 8\% for both
\bca\ and \bcf.

\begin{figure}[!htb]
\includegraphics*[width=\columnwidth]{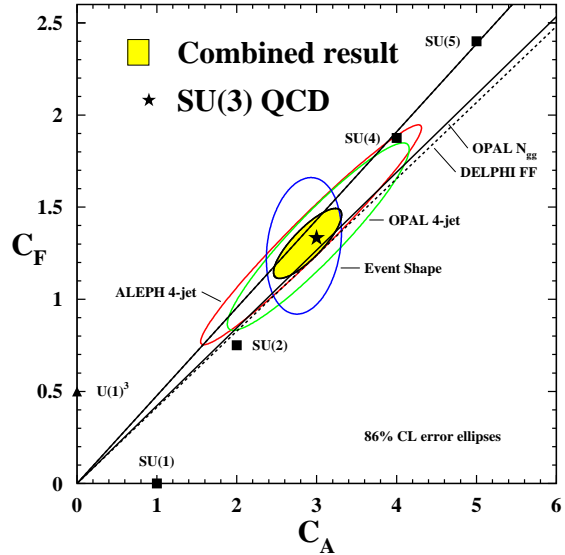}
\caption{ The figure presents various measurements of the colour
factors \ca\ and \cf\ discussed in this report. The ellipses show the
correlated measurements using 4-jet events~\cite{aleph249,OPALPR330}
or event shape distributions~\cite{colrun} while the lines represent
the results of determinations of $\ca/\cf$ from
DELPHI~\cite{delphi236} (dashed) and OPAL~\cite{OPALPR347} (solid).
The upper solid and dashed lines overlap.  The grey filled ellipsis
displays the combined result for \bca\ and \bcf\ (see text).  The
solid triangle and squares show the expectations for various
assumptions for the gauge symmetry of QCD as indicated on the
figure.  } 
\label{fig_cacf}
\end{figure}

Figure~\ref{fig_cacf} presents the results of the individual analysis
in a \cf\ vs. \ca\ plane together with the combined result and the
expectations of QCD based on the SU(3) gauge symmetry and various other
gauge symmetries.  The correlation coefficients
for~\cite{aleph249,OPALPR330} were calculated from the
references\footnote{ALEPH: $\rho=0.97$, OPAL: $\rho=0.93$.}.  The
error ellipses refer to 86\% CL.  The combined result is in good
agreement with the individual analyses and with standard SU(3) QCD
while the total uncertainties are substantially reduced.  The other
possibilities for gauge symmetries shown on the figure are clearly
ruled out.

\section{ CONCLUSIONS }

Jet physics in \epem\ annihilation based on data from $\roots=14$ to
207~GeV allows to study many aspects of QCD.  Precision measurements
of the strong coupling \as\ at many points of \roots\ gave convincing
evidence for the running of the strong coupling as predicted by the
theory. 

A combination of measurements of the mass of the b-quark \mbmz\ at the
\mz\ energy scale using jet production rates was performed.  The
resulting value
\begin{equation}
  \mbmz = (2.92\pm0.03\stat\pm0.31\syst) \mathrm{GeV}
\end{equation}
compared with low energy
measurements resulted in strong evidence for the running of the
b-quark mass analogously to the running of \as.

The investigation of the gauge structure of QCD was discussed for
several different methods: angular correlations in 4-jet final states
from hadronic \znull\ decays, global fits of event shape data at many
points of \roots, the scaling violation of the FF of gluon and quark
jets and the evolution with energy scale of the charged particle
multiplicity \nchgg\ determined from 3-jet events.  The results of the
analyses were combined taking correlations between the colour factor
measurements into account with the results:
\begin{eqnarray}
  \bca & = & 2.89\pm0.03\stat\pm0.21\syst\;\;, \\ \nonumber
  \bcf & = & 1.30\pm0.01\stat\pm0.09\syst\;\;, \\ \nonumber
  \rho & = & 0.82\;\;. \\ \nonumber
\end{eqnarray}
The combined results are in good agreement with the individual
measurements and have substantially reduced total errors of less than
10\% for both \bca\ and \bcf.  The measurements are also in good
agreement with the expectation from QCD $\ca=3$ and $\cf=4/3$.

The study of jets in QCD with \epem\ annihilation is in very good
shape and will continue to provide interesting and important new
results.


\begin{thebibliography}{10}

\bibitem{bethke92}
S.~Bethke, Z.~Kunszt, D.E. Soper, W.J. Stirling: Nucl. Phys. B {\bf 370} (1992)
  310

\bibitem{dokshitzer97b}
Yu.L. Dokshitzer, G.D. Leder, S.~Moretti, B.R. Webber: JHEP {\bf 8} (1997) 001

\bibitem{OPALPR054}
OPAL Coll., P.D. Acton et~al.: Z. Phys. C {\bf 55} (1992) 1

\bibitem{OPALPR075}
OPAL Coll., P.D. Acton et~al.: Z. Phys. C {\bf 59} (1993) 1

\bibitem{durham}
S.~Catani et~al.: Phys. Lett. B {\bf 269} (1991) 432

\bibitem{bethke00b}
S.~Bethke: MPI-PhE/2000-02 (2000)

\bibitem{pedrophd}
P.A. {Movilla Fern\'andez}: {Ph.D.} thesis, RWTH Aachen, 2003, PITHA 03/01

\bibitem{l3254}
L3 Coll., P.~Achard et~al.: Phys. Lett. B {\bf 536} (2002) 217

\bibitem{nllathmh}
S.~Catani, L.~Trentadue, G.~Turnock, B.R. Webber: Nucl. Phys. B {\bf 407}
  (1993) 3

\bibitem{nllabtbw2}
Yu.L. Dokshitzer, A.~Lucenti, G.~Marchesini, G.P. Salam: J. High Energy Phys.
  {\bf 1} (1998) 011

\bibitem{nllacp}
S.~Catani, B.R. Webber: Phys. Lett. B {\bf 427} (1998) 377

\bibitem{delphi210}
DELPHI Coll., P.~Abreu et~al.: Phys. Lett. B {\bf 456} (1999) 322

\bibitem{OPALPR299}
JADE and OPAL Coll., P.~Pfeifenschneider et~al.: Eur. Phys. J. C {\bf 17}
  (2000) 19

\bibitem{bethke00a}
S.~Bethke: J. Phys. G {\bf 26} (2000) R27

\bibitem{aleph249}
ALEPH Coll., A.~Heister et~al.: Eur. Phys. J. C {\bf 27} (2003) 1

\bibitem{delphiconf487}
DELPHI Coll., J.~Abdallah et~al.: DELPHI 2001-059 CONF 487 (2001), unpublished

\bibitem{dixon97}
L.J. Dixon, A.~Signer: Phys. Rev. D {\bf 56} (1997) 4031

\bibitem{nagy98b}
Z.~Nagy, Z.~Trocsanyi: Phys. Rev. D {\bf 59} (1998) 014020,
  Erratum-ibid.D62:099902,2000

\bibitem{rodrigo97}
G.~Rodrigo, A.~Santamaria, M.~Bilenky: Phys. Rev. Lett. {\bf 79} (1997) 193

\bibitem{bernreuther97}
W.~Bernreuther, a.~Brandenburg, P.~Uwer: Phys. Rev. Lett. {\bf 79} (1997) 189

\bibitem{nason97}
P.~Nason, C.~Oleari: Nucl. Phys. B {\bf 521} (1998) 237

\bibitem{vermaseren97}
J.A.M. Vermaseren, S.A. Larin, T.~{van Ritbergen}: Phys. Lett. B {\bf 405}
  (1997) 327

\bibitem{bilenky95}
M.S. Bilenky, G.~Rodrigo, A.~Santamaria: Nucl. Phys. B {\bf 439} (1995) 505

\bibitem{delphi177}
DELPHI Coll., P.~Abreu et~al.: Phys. Lett. B {\bf 418} (1998) 430

\bibitem{OPALPR336}
OPAL Coll., G.~Abbiendi et~al.: Eur. Phys. J. C {\bf 21} (2001) 411

\bibitem{brandenburg99}
A.~Brandenburg, P.N. Burrows, D.~Muller, N.~Oishi, P.~Uwer: Phys. Lett. B {\bf
  468} (1999) 168

\bibitem{aleph223}
ALEPH Coll., R.~Barate et~al.: Eur. Phys. J. C {\bf 18} (2000) 1

\bibitem{delphiconf644}
DELPHI Coll., P.~Bambade et~al.: DELPHI 2003-024 CONF 644 (2003), unpublished

\bibitem{el-khadra02}
A.X. El-Khadra, M.~Luke: Annu. Rev. Nucl. Part. Sci. {\bf 52} (2002) 201

\bibitem{ellis96}
R.K. Ellis, W.J. Stirling, B.R. Webber: {QCD and Collider Physics}. Vol.~8 of
  Cambridge Monographs on Particle Physics, Nuclear Physics and Cosmology,
  Cambridge University Press (1996)

\bibitem{magnoli90}
N.~Magnoli, P.~Nason, R.~Rattazzi: Phys. Lett. B {\bf 252} (1990) 271

\bibitem{nagy98a}
Z.~Nagy, Z.~Trocsanyi: Phys. Rev. D {\bf 57} (1998) 5793

\bibitem{OPALPR330}
OPAL Coll., G.~Abbiendi et~al.: Eur. Phys. J. C {\bf 20} (2001) 601

\bibitem{colrun}
S.~Kluth et~al.: Eur. Phys. J. C {\bf 21} (2001) 199

\bibitem{OPALPR134}
OPAL Coll., R.~Akers et~al.: Z. Phys. C {\bf 68} (1995) 519

\bibitem{brodsky76}
S.J. Brodsky, J.F. Gunion: Phys. Rev. Lett. {\bf 37} (1976) 402

\bibitem{konishi78}
K.~Konishi, A.~Ukawa, G.~Veneziano: Phys. Lett. B {\bf 78} (1978) 243

\bibitem{delphi236}
DELPHI Coll., P.~Abreu et~al.: Eur. Phys. J. C {\bf 13} (2000) 573

\bibitem{eden98}
P.~Eden: J. High Energy Phys. {\bf 9} (1998) 15

\bibitem{basicspqcd}
Yu.L. Dokshitzer, V.A. Khoze, A.H. Mueller, S.I. Troyan: {Basics of
  perturbative QCD}. Editions Frontieres (1991)

\bibitem{OPALPR347}
OPAL Coll., G.~Abbiendi et~al.: Eur. Phys. J. C {\bf 23} (2002) 597

\bibitem{eden99}
P.~Eden, G.~Gustafson, V.A. Khoze: Eur. Phys. J. C {\bf 11} (1999) 345

\bibitem{OPALPR038}
OPAL Coll., G.~Alexander et~al.: Phys. Lett. B {\bf 265} (1991) 462

\end{thebibliography}
\end{document}